\documentclass[apsrev,onecolumn,epsfig,tighten]{revtex4}
\usepackage[dvips]{graphicx}

\newcommand{\modulus}[1]{\vert {#1}\vert}
\newcommand{\ve}[1]{\mbox{\boldmath $#1$}}
\newcommand{\cprod}[2]{{#1}\!\times\!{#2}}

\begin{document}
\title{Vortex lattice stability and phase coherence in three-dimensional
rapidly rotating Bose condensates}

\author{S. Andrew Gifford and Gordon Baym}
\affiliation{Department of Physics, University of Illinois at
Urbana-Champaign,
 1110 West Green Street, Urbana, IL 61801\\  {\rm and}
 NORDITA, Blegdamsvej 17, DK-2100 Copenhagen \O, Denmark}
\date{\today}

\begin{abstract}

    We establish the general equations of motion for the modes of a vortex
lattice in a rapidly rotating Bose-Einstein condensate in three dimensions,
taking into account the elastic energy of the lattice and the vortex line
bending energy.  As in two dimensions, the vortex lattice supports Tkachenko
and gapped sound modes.  In contrast, in three dimensions the Tkachenko mode
frequency at long wavelengths becomes linear in the wavevector for any
propagation direction out of the transverse plane.  We compute the correlation
functions of the vortex displacements and the superfluid order parameter for a
homogeneous Bose gas of bounded extent in the axial direction.  At zero
temperature the vortex displacement correlations are convergent at large
separation, but at finite temperatures, they grow with separation.  The growth
of the vortex displacements should lead to observable melting of vortex
lattices at higher temperatures and somewhat lower particle number and faster
rotation than in current experiments.  At zero temperature a system of large
extent in the axial direction maintains long range order-parameter
correlations for large separation, but at finite temperatures the correlations
decay with separation.

\end{abstract}

\maketitle

\section{Introduction}

    Rapidly rotating Bose-Einstein condensates reveal a triangular lattice of
quantized vortices \cite{expvort,Dalibard,exp1,exp2}.  As shown analytically
in Refs.~\cite{tka,BC,MacD,Baym1,BaymT} and numerically in
Refs.~\cite{Anglin,machida,Big1}, and observed in Refs.~{\cite{exp1,exp2}} in
trapped atomic gases, the lattice supports Tkachenko modes, in which the
vortices undergo primarily transverse, but slightly elliptically polarized,
oscillations about their home positions.  Superfluid liquid helium II is
always in a regime where the Tkachenko frequencies, $\omega_T$, are linear in
the wavevector $k$ at long wavelengths.  However, by rotating the atomic
condensates sufficiently rapidly one enters a regime where the excitations in
the plane transverse to the rotation axis obey $\omega_T \propto k^2$.  As
shown in Refs.~\cite{MacD,BaymT} such soft modes in a {\it two}-dimensional
system lead to decrease of phase coherence at large separations as the lattice
rotational frequency, $\Omega_{\rm{v}}$, approaches the transverse harmonic
trap frequency, $\omega_{\perp}$, even at zero temperature, $T$.  At $T=0$,
the vortex lattice is stable and ordered until quantum fluctuations become
sufficiently large to melt the lattice; however, at finite temperature, the
lattice lacks long-range order in two dimensions \cite{BaymT}.

    In this paper we examine the stability of the vortex lattice and the long
range phase coherence in a fully three dimensional rotating BEC, taking into
account contributions of vortex line bending to the lattice modes.  This
analysis is an extension of that given in Ref.~\cite{BaymT} for excitations in
two dimensions.  We first formulate the basic equations governing the modes
for a general density profile, and then show that a uniform system has two
elasto-hydrodynamic modes:  a gapped sound mode (referrred to as an inertial
mode in \cite{Baym1,BaymT}), whose frequency is relatively isotropic; and a
Tkachenko mode, whose frequency, $\omega_T$, depends strongly on the direction
of propagation.  As we find, $\omega_T$ is quadratic at long wavelengths only
in the transverse plane.  For given $k$, $\omega_T$ becomes linear as the
wavevector $\vec k$ moves out of the transverse plane by an angle
$\delta\theta(k)$, which vanishes as $k\to 0$.

    In analyzing the implications of the modes for lattice stability and phase
coherence we assume for simplicity that the system is uniform in the axial
direction, and of thickness $Z$; we do not attempt to model realistic
condensate geometry here (as in \cite{Anglin,Big1,machida,cozzini}).  As a
consequence of the stiffening of the Tkachenko mode as the wavevector moves
out of the transverse plane, the correlations of the vortex displacements at
zero temperature remain convergent at large distances; at finite temperatures,
however, the displacement correlations, as in He-II \cite{Baym4}, grow
logarithmically with increasing separation, as long as the system is bounded
in the axial direction.  The order-parameter correlations in three dimensions
are finite at large transverse separations at $T=0$ for infinite thickness,
but fall algebraically for $Z\to 0$.  At finite temperature, the
order-parameter correlations decay algebraically with transverse separation
for infinite $Z$, and exponentially for small $Z$, as in He-II.  Analogous
issues are also encountered in the vortex lattices of superconductors
{\cite{Blatter}}.

    For sufficiently large thermally excited vortex displacements, the vortex
lattice is expected to melt.  In current experiments on rapidly rotating
Bose-Einstein condensates \cite{exp1,exp2} the vortex arrays are ordered.
However, as we argue on the basis of the Lindemann criterion, with reasonable
decreases in particle numbers and increases in rotation rates and temperature,
thermal melting of the vortex lattice could be observable.  Unlike in the case
of quantum melting of the lattice at zero temperature \cite{MacD,BaymT} where
the system makes a transition eventually to a highly correlated state,
thermally induced melting is classical in nature.  Since the system is
quasi-two dimensional, we expect the melting to be defect-mediated
\cite{Nelson}, accompanied by observable dislocations leading to a disordered
array of vortices.

\section{Elastohydrodynamic Modes}

    The basic equations describing two-dimensional motions in the system, were
derived in Ref.~\cite{BaymT} using an elasto-hydrodynamic approach.  In
constructing a full three dimensional description here we also include the
bending tension of the vortex lines and fluid velocities in the axial
direction.  As in \cite{BaymT}, we describe the deviations in the transverse
plane of the vortices from their equilibrium positions by the continuum
displacement field, $\ve{\epsilon}(\ve{r},t)$.  We assume that the vortex
lattice is rotating at an angular frequency $\Omega$, and work in the frame
co-rotating with the lattice.  To take into account the bending of the
vortices, we allow the displacement field to vary in the $z$ direction
(parallel to the rotation axis).  We ignore contributions of the normal fluid
at finite temperature here.

    As required by conservation of vorticity, the velocity in the rotating
frame, $\ve{v}$, is related to the phase, $\Phi$, of the order parameter and
the vortex displacement field, by
\begin{equation}
 \label{velophase}
 \ve{v} + 2\cprod{\ve{\Omega}}{\ve{\epsilon}} = \frac{\hbar}{m}\ve{\nabla}\Phi.
\end{equation}
The time derivative of this equation is the superfluid acceleration equation,
\begin{equation}
 \label{accequ}
 m\left(\frac{\partial\ve{v}}{\partial t} +
 2\cprod{\ve{\Omega}}{\dot{\ve{\epsilon}}}\right) =
 \hbar\ve{\nabla}\dot{\Phi} = -\ve{\nabla}\mu_{s},
\end{equation}
where $\mu_{s}$ is the chemical potential of the system.

    The equation of particle conservation is, as usual,
\begin{equation}
 \label{contequ}
 \frac{\partial n}{\partial t} + \ve{\nabla}\cdot\ve{j}  = 0,
\end{equation}
where $n(\ve{r})$ and $\ve{j}= n\ve{v}$ are the local particle density
and particle current.  Conservation of momentum in three dimensions
has the same form as in \cite{BaymT},
\begin{equation}
 \label{momcon}
   m\left(\frac{\partial\ve{j}}{\partial t} +
 2\cprod{\ve{\Omega}}{\ve{j}}\right) +
  \ve{\nabla}P = -\ve{\sigma} - \ve{\zeta},
\end{equation}
where $P$ is the pressure of the superfluid.  The elastic stress,
$\ve{\sigma}$, is given by the derivative $\delta {\cal
E}/\delta\ve{\epsilon}$ of the elastic energy with respect to the
displacement.  We include an external force, $\ve{\zeta}$, for later
convenience in calculating correlation functions.

    The total energy density of the system is,
\begin{equation}
\mathcal{E}_{\rm total}(\ve{r}) = \frac12 mn\modulus{\ve{v}}^2 +
{\cal E}_{\rm internal} + {\cal E}_{\rm el}(\ve{r}).
\end{equation}
where the pressure, ${\cal E}_{\rm internal}$, includes the effects of the
centrifugal force and external potential along with the interaction energy,
and the elastic energy, with line bending, is given by
\begin{equation}
{\cal E}_{\rm el}(\ve{r}) = 2C_1 (\ve{\nabla}\cdot\ve{\epsilon})^2
  +C_2\left[\left(\frac{\partial \epsilon_x}{\partial x}
  -\frac{\partial\epsilon_y}{\partial y}\right)^2 + \left(\frac{\partial
  \epsilon_x}{\partial y} +\frac{\partial \epsilon_y}{\partial
 x}\right)^2\right] + \frac{\tau}2\left\vert\frac{\partial
 \ve{\epsilon}}{\partial z}\right\vert^2 +
  \gamma n_v^0 \ve\epsilon \cdot \ve{\nabla}_{\bot}n.
 \label{elastic}
\end{equation}
Here $C_1$ and $C_2$ are respectively the compressional and shear moduli
of the vortex lattice, and $\tau$ is the vortex line bending tension.  The
final term describes the coupling of the particle and vortex densities
\cite{elastic}; the coupling constant $\gamma$ is a function of rotation rate.
In an incompressible fluid, $C_2
= n\hbar\Omega/8 = -C_1$ and $\gamma=(\pi\hbar^2/m)\ln(\ell/\xi_{\rm c})$, where
$\ell$ and $\xi_{\rm c}$ are the radii of the Wigner-Seitz cell and core of
any given vortex, respectively; in the limit where only the lowest Landau
level (LLL) in the Coriolis force is occupied ($\Omega\to\omega_\perp$), $C_1
= 0$, $C_2 \approx 0.1191ms^2 n$ and $\gamma=\pi\hbar^2/m$
\cite{sonin,elastic}.  For vortex core size, $\xi$, small compared with the
spacing between vortices,
\begin{equation}
 \tau = -\frac{1}{2}\hbar\Omega n\ln (\Omega/\Omega_{\rm C }),
\label{tau}
\end{equation}
where $\Omega_{\rm C} \sim \hbar/m\xi^2 > \Omega$.  More generally, we can
estimate $\tau$ from the local energy of the vortex within a unit cell, $\sim
|\hbar\nabla f(r_\perp)|^2/2m$, where $f(r_\perp)$ is the order parameter
within the cell, normalized so that its average over a unit cell is unity.
From Ref.~\cite{baym2}, we have,
\begin{equation}
   \tau = n\hbar\Omega a(\Omega),
  \label{taua}
\end{equation}
where
\begin{equation}
  a= \int \left\{\left( \frac{\partial f}{\partial\rho}\right)^2 +
  \frac{f^2}{\rho^2}\right\}\frac{d^2\rho }{2\pi},
\end{equation}
where $\rho$ is the transverse radius measured from the center of the
cell, and the integration is over the cell.  For small rotation,
Eq.~(\ref{taua}) agrees with Eq.~(\ref{tau}), while in the LLL limit, $a\to
1.7$ \cite{baym2}.

    From Eq.~(\ref{elastic}) we find,
\begin{equation}
 \label{stress}
 \ve{\sigma} = -4\ve{\nabla}_{\bot}(C_{1}\ve{\nabla}_{\bot}\cdot\ve{\epsilon})
   - 2(\ve\nabla_{\bot}\cdot C_{2}\ve\nabla_{\bot})\ve{\epsilon}
   - \frac{\partial}{\partial z}\left(\tau\frac{\partial\ve{\epsilon}}{\partial z}\right)
   + \gamma n_v^0 \ve{\nabla}_{\perp}n.
\end{equation}
If the density of the system is non-uniform the gradients act on the
elastic constants as well.  The elasto-hydrodynamic modes of the system follow
from the superfluid acceleration equation and conservation laws of momentum
and particle number given above.  These equations are valid for general
density dependence in equilibrium.

    We are interested in the small variations of wavevector $\ve{k}$ and
frequency $\omega$, in the velocity, displacement field, and density.  We
obtain two coupled equations.  Taking the Fourier transform of
Eq.~({\ref{mainequ1}}), we have,
\begin{equation}
 \label{ftmainequ1}
 \left(\omega^{2} - s^{2}k^{2} - \frac{\hbar^2k^4}{4m^2}\right)\delta n
  +\left(4n\Omega^{2} + \frac{\gamma n_{\rm
 v}^0nk^2}{m}\right)i\ve{k}\cdot\ve{\epsilon} =
 -\frac{i}{m}\ve{k}\cdot(\ve{\sigma} + \ve{\zeta}),
\end{equation}
where $i\ve{k}\cdot\ve{\sigma} =
((4C_1+2C_2)k_{\bot}^2 + \tau k_z^2)i\ve{k}\cdot\ve{\epsilon} -
\gamma n_{\rm v}^0k^2\delta n$; Fourier transforming Eq.~(\ref{mainequ2}),
and using the curl of Eq.~({\ref{velophase}}) for $v_{z}$ together with
the equation of continuity, we have
\begin{equation}
 \label{ftmainequ2}
 \left(\omega^{2} - \frac{2C_{2}}{nm}k_{\bot}^{2} - \frac{\tau}{nm}k_{z}^{2} -
 4\Omega^{2}\frac{k_{z}^{2}}{k^{2}}\right)i\ve{k}\cdot\ve{\epsilon}
 + \frac{\omega^{2}k_{\bot}^{2}}{nk^{2}}\delta n =
 \frac{ik_{z}^{2}}{mnk^{2}}\ve{k}\cdot(\ve{\sigma} + \ve{\zeta}).
\end{equation}
From Eq.~(\ref{divdiff1}), the longitudinal and transverse components,
$\ve{\hat{k}}\cdot\ve{\epsilon}
\equiv \epsilon_{L} $ and $ (\cprod{\ve{\hat{k}}}{\ve{\epsilon})_{z}} \equiv
\epsilon_{T} $, respectively, are related by,
\begin{equation}
 \label{ftdivdiff1}
  i\omega\left(1 - \frac{\gamma n_{\rm v}^0 k^2}{\pi
 m(\omega^2-s'^2k^2)}\right)(\cprod{\ve{\hat{k}}}{{\ve{\epsilon}})_{z}} =
 2\Omega\left(1 + \frac{2(C_2+2C_1)k^2_{\bot} + \tau k^2_z}{4\Omega^2mn} +
 \frac{(\gamma n_{\rm v}^0)^2 k^4}{4\Omega^2m^2(\omega^2-s'^2k^2)}
 \right)\ve{\hat{k}}\cdot\ve{\epsilon},
\end{equation}
where we have the modified speed of sound, $s'^2 = s^2 + \hbar^2k^2/4m^2$.
For $k\to 0$, we have $i\omega(\cprod{\ve{\hat{k}}}{{\ve{\epsilon}})_{z}} =
2\Omega\ve{\hat{k}}\cdot\ve{\epsilon}$.

    Equations~({\ref{ftmainequ1}}) and (\ref{ftmainequ2}) give the frequencies
and mode functions for the rapidly rotating BEC; eliminating $\delta n$ and
$\ve{\hat{k}}\cdot\ve{\epsilon}$ we find the secular equation for the mode
frequencies,
\begin{eqnarray}
 D(\ve{k},\omega) \equiv & \omega^4 - \alpha\omega^2 + \beta
  = (\omega^2 - \omega_I^2)(\omega^2- \omega_T^2)= 0,
\label{bigD1}
\end{eqnarray}
with
\begin{eqnarray}
 \alpha & = &
  4\Omega^{2} + s'^{2}k^2 +
  \frac{(4C_1+4C_2)k_{\bot}^{2}}{mn} 
  + \frac{2\tau}{mn}k_{z}^{2} + \frac{2\gamma n_v^0}{m}k_{\perp}^2,
  \\
 \beta & = & s'^{2}\left(4\Omega^{2}k_{z}^{2} +
  \frac{2k_{\bot}^{2}}{mn}(2C_{1}k_{z}^{2} + C_{2}(k_{\perp}^2 + 2k_z^2)) +
  \frac{\tau k_{z}^{2}}{mn}(k_{\perp}^2 + 2k_z^2)\right)
  + \left(\frac{\gamma n_{\rm v}^0}{m}\right)^2k_{\perp}^2k_z^2.
  \nonumber
\end{eqnarray}

The second term in $\beta$ is only significant for $\hbar\Omega>>ms^2$.  
The mode with frequency $\omega_{I}$ is the gapped sound mode, and with frequency
$\omega_{T}$, the Tkachenko mode.  In the soft limit $\hbar\Omega\gg sk$,
$\alpha^2 \gg 4\beta$ and we find, $\omega_I^2 \approx \alpha$ and
$\omega_{T}^{2} \approx \beta/\alpha$.  Explicitly, in the incompressible limit,
\begin{equation}
\label{tkachf-inc}
 \omega_{T}^{2} \approx \frac{s^{2}\left(4\Omega^{2}k_{z}^{2} +
 \hbar\Omega\left(k_{\bot}^{4} -
  2k_{z}^{2}(k_{\bot}^2 +
   2k_z^2)\ln\left(\Omega/\Omega_{\rm c}\right)\right)/4m\right) + (\hbar\Omega\ln(\Omega/\Omega_{\rm c})/2m)k_{\perp}^2k_z^2}{4\Omega^{2} + s^2k^2},
\end{equation}
and in the LLL limit,
\begin{equation}
\label{tkachf-lll}
 \omega_{T}^{2} \approx \frac{\left(\hbar\Omega/m\right)^2k_{\perp}^2k_z^2 +
 s^{2}\left(4\Omega^{2}k_{z}^{2} + 1.7(\hbar\Omega k_z^2/m)\left(k_{\bot}^2 +
 2k_z^2\right)\right) + 0.24s^4k_{\bot}^2(k_{\bot}^2+2k_z^2)}{4\Omega^2 +
 (\hbar\Omega/m)\left(2k_{\bot}^2 + 3.4k_z^2\right)},
\end{equation}
where we have used the values of $C_1$, $C_2$, $\tau$, and $\gamma$ in the
LLL limit.

\begin{figure}[htbp]
\begin{center}\vspace{0.0cm}
\rotatebox{0}{\hspace{-0.cm}
{\includegraphics[width=5.5in]{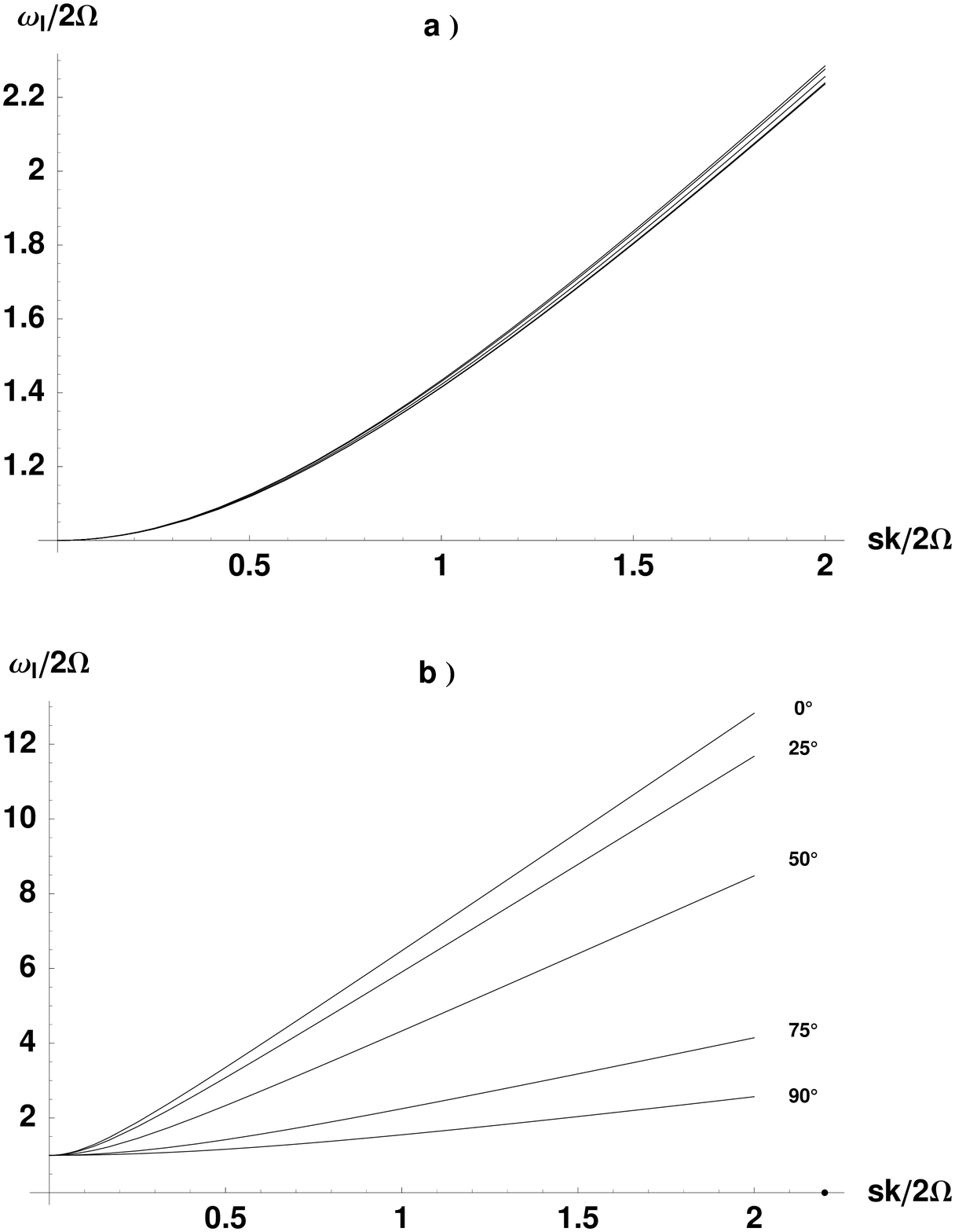}}}
\caption{
Dispersion relation for the gapped sound mode in three dimensions.  The frequency
of the mode does not change significantly with direction in the incompressible limit
(a), while in the lowest Landau limit there is a noticeable anisotropy (b).\label{fig:gsm}}
\end{center}
\end{figure}

\begin{figure}[htbp]
\begin{center}\vspace{0.0cm}
\rotatebox{0}{\hspace{0.0cm}
{\includegraphics[width=5.5in]{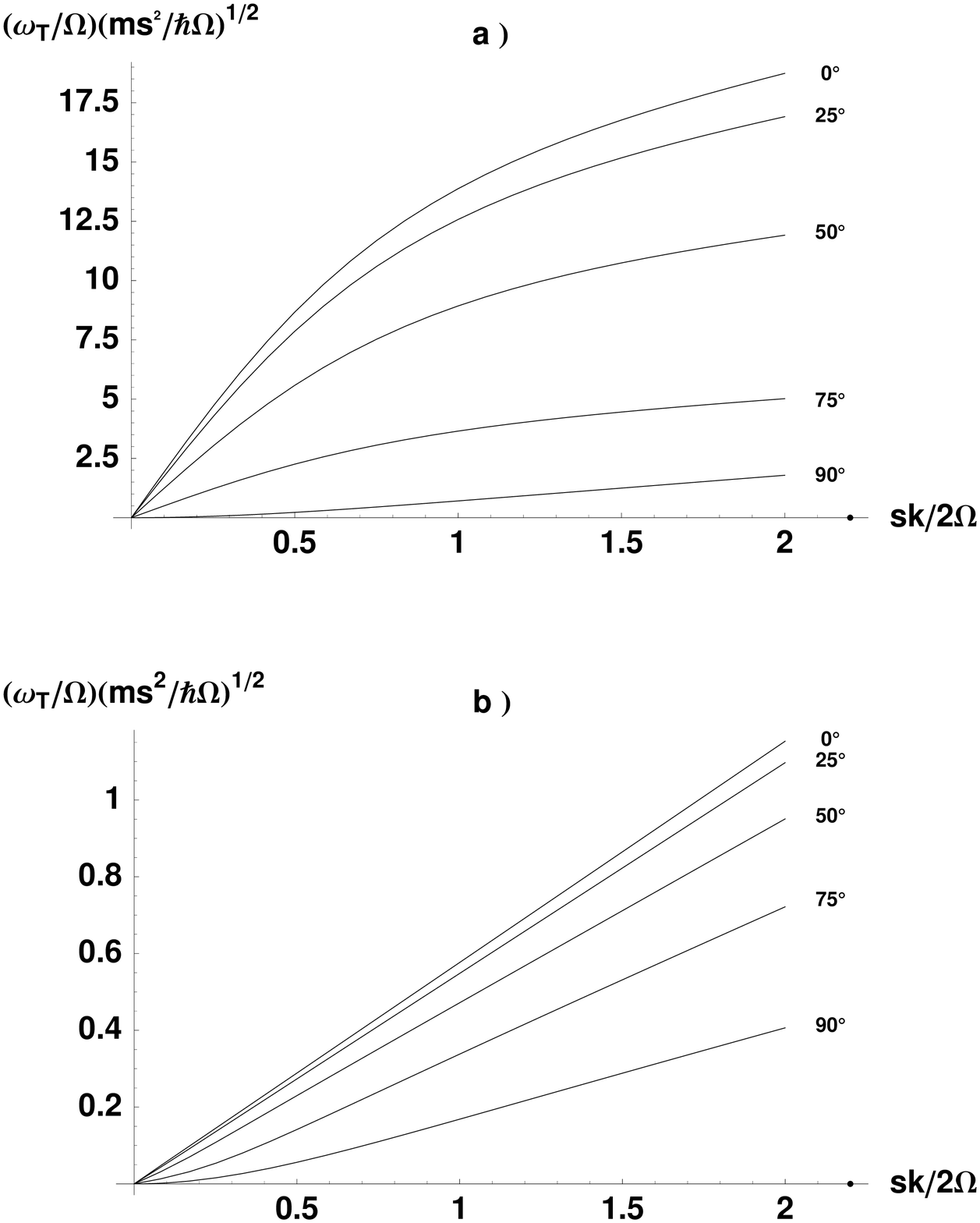}}}
\caption{
Dispersion relations for the Tkachenko mode in three dimensions.  The curves
are labelled by the azimuthal angle of the wavevector, measured from the
rotation axis.  Panel (a) shows the characteristic structure in the regime where
$\hbar\Omega$ is small compared with $ms^2$.  The parameters are
taken for a cloud of $^{87}$Rb with density $10^{14}$cm$^{-3}$ and
$\Omega = 0.9(8.3\times 2\pi)$ Hz, for which $\hbar\Omega/ms^2 \simeq 0.01$.  The
line bending tension is described by Eq.~(\ref{tau}).  Panel (b) corresponds to the
regime where $\hbar\Omega$ is large compared with $ms^2$.  The graph is calculated
for a density of $10^{11}$cm$^{-3}$ and $\Omega = 0.99(8.3\times2\pi)$ Hz so
that $\hbar\Omega/ms^2 \simeq 12$, with $\tau = 1.7n\hbar\Omega$.  The lattice Debye
wavevector corresponds to $sk/2\Omega=1/\sqrt2$; thus the elasto-hydrodynamic
description is valid only for $(sk/2\Omega)\sin
\theta$ well to the left of this point.\label{fig:tka3D}}
\end{center}
\end{figure}

    Figures~\ref{fig:gsm} and~\ref{fig:tka3D} display the gapped sound mode
frequency and the Tkachenko mode frequency, respectively, as a function of the
wavenumber, $k$.  In the plots, we use parameters consistent with the
experiments of Refs.~\cite{exp1, exp2} in $^{87}$Rb:  $a_{s}=4.76$ nm and
$\omega_{\bot}=8.3\times 2\pi$ Hz.

    From Figure~\ref{fig:gsm}, frequencies of the gapped sound mode do not
vary significantly with $\theta$ in the incompressible limit shown in
Figure~\ref{fig:gsm}a, since the elastic contributions to this mode are small
relative to the hydrodynamic terms.  However, in the lowest Landau limit, the
mode frequency does change with direction, Figure~\ref{fig:gsm}b, with the
different values of the azimuthal angle, $\theta$, in terms of which $k_{\bot}
= k\sin\theta $ and $ k_{z} = k\cos\theta $. Since $\hbar\Omega>>ms^2$ in this
limit, the line tension is the dominant.  Without rotation, $\omega_I$ would
be a sound mode; the Coriolis force introduces a gap into the dispersion.

As one sees from Fig.~\ref{fig:tka3D}, the form of the Tkachenko mode is quite
anisotropic.  For given $k$, $\omega_T(\ve{k})$ decreases with increasing $\theta$;
the Tkachenko modes have lowest energy in the transverse plane
($\theta = 90^{\circ} $).  Figure~\ref{fig:tka3D}a corresponds to the more slowly
rotating regime (a), where $\hbar\Omega/ms^2$ is small, and Fig.~\ref{fig:tka3D}b to
the rapidly rotating regime (b), where $\hbar\Omega/ms^2$ is large.  The present
elasto-hydrodynamic description of the modes is valid only for the transverse
component of the wavevector small compared with the inverse lattice spacing,
corresponding in Fig.~\ref{fig:tka3D}b, drawn for $\hbar\Omega/ms^2 \approx 12$, to
$(sk/2\Omega)\cos\theta$ well below $1/\sqrt2$.

    In the incompressible regime, the Tkachenko mode frequency is always
linear for sufficiently small $k$ out of the plane:  $\omega_T = sk\cos\theta
+ \dots$.  Only exactly in the transverse plane does the Tkachenko mode start
quadratically \cite{Sandro}.  A crossover from linear to quadratic, due to a
competition between the hydrodynamic (first) and elastic (second) terms in the
numerator of Eq.~(\ref{tkachf-inc}), occurs for $k^2 \gtrsim
(16m\Omega/\hbar)(\cos^2\theta/\sin^4\theta)$.

    In the LLL regime, the Tkackenko mode is also linear for sufficiently
small $k$ out of the transverse plane.  In the plane, the modes is purely
quadratic, and it is determined by the sound velocity.  For modes out of the
transverse plane, there is another crossover from linear to quadratic for $k
\sim 1.2ms/\hbar$.  This crossover is due the competition of the coupling of the particle and vortex densities with the line tension terms with the
product of the compression and hydrodynamic terms.

    For $k_{z}=0$, we obtain the two dimensional results given in
{\cite{Baym1, BaymT}}.  In the extreme incompressible limit, $s\to\infty$, we
obtain the expression for $\omega_{T}$ at arbitrary angles given in
{\cite{BC}}.  Furthermore, for small wavelengths in this regime, the
dispersion relation $\omega_T=2\Omega\modulus{\ve{\hat{k}}_z}$ becomes that of the
inertial mode of a rotating classical fluid {\cite{Green}}.

    Experiments have observed Tkachenko modes in rapidly rotating BECs
\cite{exp1}.  These Tkachenko modes are excited by creating a hole in the
center of the condensate with a laser; the Coriolis force on the surrounding
superfluid shears the vortex lattice.  The clouds in these experiments are
relatively flat, with the Tkachenko modes confined to two dimensions.  The
Tkachenko mode is approximately modeled from the homogeneous spectrum of
Eq.~(\ref{tkachf-inc}) for $k_z=0$ \cite{Baym1}.  The use of a co-rotating
optical lattice to manipulate the structure of the vortex lattice provides
another avenue to exciting Tkachenko modes \cite{cornelloptical}.  One can
imagine using a co-rotating ``optical annulus" to tug on vortices about the
center of a rapidly rotating BEC.  By tilting the annulus, one can excite
modes out of the plane ($k_z\ne 0$).  To observe the out of plane modes, the
system sizes along the azimuthal axis and in plane ($R_{{\rm TF},\bot}>>\ell$)
should be comparable.

\section{Correlation Functions}

    From Eqs.~(\ref{ftmainequ1}) and (\ref{ftmainequ2}) for the lattice
displacement field and density variation, we derive the correlation functions
for the vortex displacement field and the phase.  In general, the correlation
functions of the lattice displacements are given by
$\left<\epsilon_{i}\epsilon_{j}\right>(\ve{k},\omega) =
\delta\left<\epsilon_{i}\right>/\delta\zeta_{j}$, where the indices denote the
transverse and longitudinal components.  For a rapidly rotating Bose gas,
these correlation functions are,
\begin{equation}
 \label{LLcorr}
 \left< \epsilon_{L}\epsilon_{L}\right> = \frac{\omega^{2} -
 s^{2}k^{2}}{nmD(\ve{k},\omega)}\left(\frac{\omega^{2} -
 s^{2}k_{z}^{2}}{\omega^{2} - s^{2}k^{2}} + \frac{2C_2 k_{\bot}^{2}
 + \tau k_{z}^{2}}{4\Omega^2 nm}\right),
\end{equation}
\begin{equation}
 \label{TTcorr}
 \left< \epsilon_{T}\epsilon_{T}\right> = \frac{\omega^{2} -
 s^{2}k^{2}}{nmD(\ve{k},\omega)}\left(1 +
 \frac{2(2C_{1} + C_{2})k_{\bot}^{2} +
 \tau k_{z}^{2}}{4\Omega^2 nm} - 
 \frac{\gamma^2 n_v^2 k^2 k_{\perp}^2}{4m^2\Omega^2(\omega^2-s^2k^2)}\right),
\end{equation}
and
\begin{equation}
\left< \epsilon_{L}\epsilon_{T}\right> =
\left<\epsilon_{T}\epsilon_{L}\right>^* =\frac{i\omega(\omega^{2} -
 s^{2}k^{2})}{2\Omega nmD(\ve{k},\omega)}\left(1 - \frac{\gamma n_{\rm v}^{0}k_{\perp}^2}{m(\omega^2-s^2k^2)} \right).
\end{equation}
where $D(\ve{k},\omega)$ is defined in Eq.~(\ref{bigD1}).  The coupling of the particle and
vortex densities only comes into play in the transverse-longitudinal correlation
functions.  These off-diagonal terms are not important for calculating the real space
elastic displacement correlations.

We consider a system bounded between two planes perpendicular to the rotational
axis with periodic boundary conditions in the $z$ direction; $Z$ is the distance of
separation of the two planes.  In the limit, $\Omega \gtrsim s/{\cal R}_\perp$, where
${\cal R}_\perp$ is the transverse size of the system, the real space correlation of the vortex displacements is
\begin{equation}
 \label{vorcorrm}
  \left<\modulus{\ve{\epsilon}(\ve{r}\,)-\ve{\epsilon}(\ve{r}\,')}^{2}\right> \approx \frac{\hbar}{Z}\sum_{k_{z}} \int \frac{1 - \cos \ve{k}\cdot\ve{R}}{(\omega_{I}^{2}-\omega_{T}^{2})mn}
  \left[\frac{s^{2}k_{\bot}^{2}}{\omega_{T}}(1+2f(\omega_{T}))\right]\frac{d^{2}k_{\bot}}{(2\pi)^{2}}
  + {\rm regular~terms}.
\end{equation}
where $\ve{R} = \ve{r} - \ve{r}\,' = (\ve{R}_{\bot},R_{z})$, and $f(\omega)$ is the
Bose distribution function, $f(\omega) = (e^{\beta\hbar\omega} - 1)^{-1}$.  The
$k_z=0$ term is the two dimensional result; the displacement correlation function
in three dimensions is augmented by the $k_{z} \ne 0$ terms in the sum.  At $T=0$,
the right side of Eq.~({\ref{vorcorrm}}) is convergent.  However, depending on the
geometry, the softness of the Tkachenko mode, i.e., $\omega_T\propto k^2_\perp$,
can lead to an infrared divergence as $|\ve{R}_\perp|\to\infty$ of the finite temperature
part of the first term on the right side of Eq.~({\ref{vorcorrm}}).  The regular terms
include contributions from the gapped sound mode which, because of the gap, do not diverge.

Extracting the most divergent parts of the correlation function, taking the leading
terms in the Tkachenko mode frequencies, we find $\modulus{\ve{R}_{\bot}} \gg \ell$,
\begin{equation}
 \label{vorcorrT}
 \frac{\left<\modulus{\ve{\epsilon\,}(\ve{r}\,)-\ve{\epsilon\,}(\ve{r}\,')}^{2}\right>}{\ell^2}
 \approx\frac{Tm\Omega}{4\pi\hbar C_{2}Z}\ln\left[\frac{\sinh(Z/\ell)}{\sinh(Z\ell/4\modulus{\ve{R}_{\bot}}^2)}\right],
\end{equation}
where $\ell = (\pi n_{\rm v})^{-1/2}$ is the radius of the Wigner-Seitz cell of each
vortex.  If $Z$ is finite, then the displacement correlations diverge when
$\modulus{\ve{R}_{\bot}} \to \infty$.  On the other hand, for $Z \ll {\cal R}_\perp^2/\ell$,
the right side of Eq.~({\ref{vorcorrT}}) is proportional to $\ln N_{\rm v}(\ve{R}_\bot)$,
where $N_{\rm v}(\modulus{\ve{R}_{\bot}})= (\modulus{\ve{R}_{\bot}}/\ell)^2$ is the
number of vortices within radius $R_\perp$ -- the result for two dimensions {\cite{BaymT}}.
In this limit, no motion is excited in the $z$ direction.  For
$Z \sim \modulus{\ve{R}_{\bot}} \to \infty $, the displacement correlations become
independent of $\modulus{\ve{R}_{\bot}}$ and $Z$, and the vortex lattice maintains
long range order, with the vortices oscillating a finite degree about their equilibrium
positions, due to quantum and thermal fluctuations.

    When the vortex displacements become sufficiently large, due to quantum
and thermal fluctuations, the lattice is expected to melt.  To determine the
mean square displacements of the vortices in a realistic trapping geometry
requires solving Eqs.~(\ref{velophase})-(\ref{momcon}) and (\ref{stress}),
taking the spatially varying density into account.  As a first estimate, we
employ the dispersion relation for the modes in uniform geometry, evaluated at
the center of the trap \cite{Fischer}, and find
\begin{equation}
  \frac{\left<\modulus{\ve{\epsilon}}^2\right>}{\ell^2} \sim
     \Gamma\left[N\left(1-\frac{\Omega^2}{\omega_\perp^2}\right)\right]^{-4/15}
 \left(\frac{T}{T_{\rm c}}\right)\ln[N_{\rm v}({\cal R}_\perp)],
 \label{eps2}
\end{equation}
where
\begin{equation}
 \Gamma =
 \frac{8}{15^{3/5}\zeta(3)^\frac13}\left(\frac{a_s
  b}{d_\perp}\right)^{2/5}\left(\frac{\omega_z}{\omega_\perp}\right)^{11/15};
\end{equation}
here $d_\perp$ is the transverse oscillator length, $b\gtrsim 1$ is the
Abrikosov parameter, describing the renormalization of $g$ arising from
density variations within the vortex cells \cite{MacD,Baym1,baym2}, we have
taken $C_2=n\hbar\Omega/8$; and $T_c\approx
\hbar\left(N\omega_z(\omega_\perp^2-\Omega_{\rm v}^2\right)/\zeta(3))^{1/3}$
is the Bose-Einstein condensation temperature \cite{String}.

    For the JILA experiments \cite{exp1,exp2}, $\Gamma\simeq 0.07$.  Thus the
Lindemann criterion for melting of the vortex lattice \cite{RS},
$\langle\modulus{\ve{\epsilon}}^2\rangle/\ell^2\simeq 0.07$, would indicate that the
currently observed lattices should be ordered, as observed, but at higher
temperatures, and somewhat lower particle number and faster rotation rates,
one could observe thermal melting of the vortex lattice.  For example, with
$N\simeq 10^4$ and $\Omega/\omega_\perp \simeq 0.999$, Eq.~(\ref{eps2}) gives
$\langle\modulus{\ve{\epsilon}}^2\rangle/\ell^2\sim 0.2T/T_c$, so that one is in the
range of melting.  The drop in $C_2$ in the LLL region, which is not included
in this estimate, would further increase the vortex displacements.  In the LLL
limit, quantum fluctuations contribute to the vortex displacements an amount
$\left<\vert\ve{\epsilon}\vert^2\right>/\ell^2 \approx 1.22/\nu$, where $\nu =
N/N_{\rm v}$ is the filling factor \cite{BaymT,MacD}.  The ratio of the mean
square displacements due to thermal fluctuations to those due to quantum
fluctuations is
\begin{equation}
\label{qtf_ratio}
\frac{\left<\modulus{\ve{\epsilon}}^2\right>_{\rm thermal}}{\left<\modulus{\ve{\epsilon}}^2\right>_{\rm quantum}}\sim 10^3\nu^{-{2/3}}(T/\hbar\Omega)\ln N_{\rm v}.
\end{equation}
As $\nu$ decreases, this ratio grows, and eventually thermal fluctuations,
at a given temperature, will dominate over quantum fluctuations.

    The Lindemann criterion gives only a rough approximation for determining
the point of vortex lattice melting.  Correlation functions of the type in
(\ref{vorcorrT}) with logarithmic terms are characteristic of two dimensional
systems that exhibit a Kosterlitz-Thouless transition, suggesting that the
thermal melting of the vortex lattice is defect-mediated {\cite{Nelson}}.
Experimental determination of the breakdown of the vortex lattice order, and
the detailed conditions under which the lattice melts, would provide valuable
insight.  We discuss this theoretical avenue in a future article
\cite{defmelt}.

    We now derive the phase-phase correlation function, which we write in
terms of displacment and velocity correlations using the Fourier transform of
Eq.~({\ref{velophase}}) as \cite{BaymT}:
\begin{eqnarray}
 \left<\Phi\Phi^*\right> & = & \frac{m^{2}}{\hbar^2 k^{2}}\left[\left<v_{L}v_{L}^*\right> -
 2\Omega(\left<\epsilon_{T}v_{L}^*\right> +
 \left<v_{L}\epsilon_{T}^*\right>) +
 4\Omega^{2}\left<\epsilon_{T}\epsilon_{T}^*\right>\right].
\end{eqnarray}
The density-density correlation function,
\begin{equation}
 \label{dencorr}
  \left< \delta n\delta n^*\right> =
 \frac{nk^{2}}{mD(\ve{k},\omega)}\left[\omega^{2} 
 - 4\Omega\frac{k_z^2}{k^2}
 - \frac{2C_{2}}{nm}k_{\bot}^{2}\left(1+\frac{k_z^2}{k^2}\right) 
 - \frac{4C_1k_{\perp}^2}{nm}\left(\frac{k_z^2}{k^2}\right)
 - \frac{\tau k_{z}^2}{nm}\left(1+\frac{k_z^2}{k^2}\right)\right],
\end{equation}
together with the equation of continuity, gives
\begin{eqnarray}
 \label{vlvlcorr}
 \left< v_{L}v_{L}^*\right> &=& \frac{\omega^{2}}{n^{2}k^{2}}\left< \delta
 n\delta n^*\right> - \frac{1}{nm}
 \\
 &=&
 \frac{\omega^{2}}{nmD(\ve{k},\omega)}\left[\omega^{2} 
 - 4\Omega\frac{k_z^2}{k^2}
 - \frac{2C_{2}}{nm}k_{\bot}^{2}\left(1+\frac{k_z^2}{k^2}\right) 
 - \frac{4C_1k_{\perp}^2}{nm}\left(\frac{k_z^2}{k^2}\right)
 - \frac{\tau k_{z}^2}{nm}\left(1+\frac{k_z^2}{k^2}\right)\right] -
 \frac{1}{nm},
\end{eqnarray}
where the constant term comes from the commutation relation between the
current density and the particle density.  Furthermore, using
Eq.~(\ref{densityelastic}) and the equation of continuity,
\begin{eqnarray}
 \label{vleTcorr}
 \left< v_{L}\epsilon_{T}^*\right> &=& \frac{2\Omega\omega^{2}}{\omega^{2} -
 s^{2}k^{2}}\left<\epsilon_{T}\epsilon_{T}\right> -
 i\frac{\gamma n_{\rm v}^0\omega k^2}{m(\omega^2-s^2k^2)}\left<
 \epsilon_{L}\epsilon_{T}\right>
 \\
 \left< \epsilon_{T}v_{L}^*\right> &=& \frac{2\Omega\omega^{2}}{\omega^{2} -
 s^{2}k^{2}}\left<\epsilon_{T}\epsilon_{T}\right> +
 i\frac{\gamma n_{\rm v}^0\omega k^2}{m(\omega^2-s^2k^2)}\left<
 \epsilon_{T}\epsilon_{L}\right>
\end{eqnarray}
We thus find the correlation function of the phase:
\begin{eqnarray}
 \left<\Phi\Phi^*\right>
  & = &\frac{m}{n\hbar^2k^{2}D(\ve{k},\omega)}\bigg\{\left(s^{2}k^{2}
  - 4\Omega^2\frac{k_{z}^2}{k^2}
  - \frac{(4C_1+2C_2)k_{\bot}^2 + \tau k_{z}^2}{nm}\left(\frac{k_z^2}{k^2}\right)
  - \left(\frac{\gamma n_{\rm v}^0}{m}\right)^2\frac{k^2k_{\perp}^2}{\omega^2-s^2k^2}
 \right)\omega^{2}
 \\
 &-& s^{2}\left(\left(4\Omega^{2} +  \frac{4C_1k_{\bot}^2}{nm}\right)(k_{\bot}^2 + 2k_{z}^2) +
 \left(\frac{2C_2k_{\bot}^2}{nm} + \frac{\tau k_z^2}{nm}\right)(2k_{\bot}^2+3k_z^2)\right)
 - \left(\frac{\gamma n_{\rm v}^0}{m}\right)^2\left(k_{\perp}^2k_z^2 + \frac{k^2k_{\perp}^2}{\omega^2-s^2k^2}\right)
 \nonumber.
\end{eqnarray}
In the soft limit, the dominant term in the position space phase-phase
correlation is
\begin{equation}
 \label{rsphcorr1}
  \left< \vert\Phi(\ve{r}\,)-\Phi(\ve{r}\,')\vert^{2}\right> \approx
   \frac{1}{Z}\sum_{k_{z}}\int \frac{(1-\cos\ve{k}\cdot\ve{R})}{(\omega_{I}^2
 -\omega_{T}^2)}\frac{4\Omega^{2}ms^2}{n\hbar\omega_{T}}
 \left(1+\frac{k_{z}^{2}}{k^{2}}\right)
 (1+2f(\omega_{T}))\frac{d^{2}k}{(2\pi)^{2}} +
 {\rm regular~terms}.
\end{equation}

    At $T=0$, this correlation function is convergent as $ Z \to \infty $, but
for $Z \to 0$ it grows logarithmically with increasing separation
{\cite{BaymT}}.  At finite temperatures, for $\vert\ve{R}\vert\gg\ell$,
\begin{equation}
 \label{Trsphcorr1}
  \left< \vert\Phi(\ve{r}\,)-\Phi(\ve{r}\,')\vert^{2}\right> \approx
  \frac{T}{\pi\hbar^2}\sqrt{\frac{2\Omega^2 m^3}{C_2
    n}}\int^{k_D}_{\Lambda^{-1}} \coth{\left(Z\sqrt{\frac{C_2}{2\Omega^2
 nm}}k_\perp^2\right)}\frac{dk_\perp}{k_\perp} +
{\cal{O}}(k_D\Lambda)^{-\frac12},
\end{equation}
where $k_D^2=4\pi n_{\rm v}$ is the Debye wavelength, cutting off modes with
wavelengths less than the lattice spacing, and
\begin{equation}
  \Lambda^2 = \vert\ve{R}_{\bot}\vert^{2} + R_{z}\sqrt{C_{2}/2\Omega^{2}nm}
\end{equation}
is the square of the effective infrared cutoff.  For $Z \gg {\cal
R}_{\perp}^2/\ell$, Eq.~({\ref{Trsphcorr1}}) increases logarithmically as
$\vert\ve{R}\vert\to\infty$:
\begin{equation}
  \label{3Dpp}
  \left<\vert\Phi(\ve{r})-\Phi(\ve{r}')\vert^{2}\right> \approx
  \frac\eta2\ln\left(k_D^2\vert\ve{R}_{\bot}\vert^{2} +
  k_D^2 R_{z} \sqrt{C_{2}/2\Omega^{2}nm}\,\right),
\end{equation}
where $\eta = (T/\pi\hbar^2)\sqrt{\Omega^2 m^{3}/2C_{2}n}$.  If we assume
that the fluctuations of the order parameter, $ \Psi $, are Gaussianly
distributed, then
\begin{equation}
  \left<\delta\Psi(\ve{r})\delta\Psi(\ve{r}')\right> \sim e^{-\frac{1}{2}\left<\vert\Phi(\ve{r})-
   \Phi^{*}(\ve{r}')\vert^{2}\right>} \sim (k_D\Lambda)^{-2\eta}.
\end{equation}
On the other hand, for $Z\ll \hbar/ms$, Eq.~(\ref{Trsphcorr1}) gives
\begin{equation}
  \label{2Dpp}
  \left< \vert\Phi(\ve{r})-\Phi(\ve{r}')\vert^{2}\right> \approx \frac{2\Omega^2 m^2 T}{3\pi\hbar^2 C_{2}Z}\Lambda^2,
\end{equation}
which implies an exponential loss of long range order as $\modulus{\ve{R}}\to\infty$.
Thus, thermal fluctuations can have a stronger effect on destroying phase coherence
than quantum fluctuations, which depend only logarithmically on $\Lambda$.  For
finite temperatures in the LLL limit, the ratio of the contribution to
$\left<\modulus{\Phi(\ve{r})-\Phi(\ve{r}')}^{2}\right>$ from thermal fluctuations to the
contribution from quantum fluctuations is
$\sim 10^3\nu^{-\frac23}(T/\hbar\Omega)N_{\rm v}(|\ve{R}_{\perp}|)/\ln N_{\rm v}(|\ve{R}_{\perp}|)$,
which increases with $|\ve{R}_\perp|$.

\vspace{12pt}

    This work was supported in part by NSF Grants PHY00-98353, PHY03-55014,
PHY05-00914, and PHY07-01611.  We thank NORDITA for hospitality while this
work was completed.

\section*{APPENDIX}

    In this Appendix, we derive Eqs.~(\ref{ftmainequ1})-({\ref{ftdivdiff1}}).
Dividing Eq.~({\ref{momcon}}) by $n$, using the Gibbs-Duhem relation at $T=0$,
$\ve{\nabla}P = n\ve{\nabla}\mu_{s}$, and subtracting it from
Eq.\,({\ref{accequ}}), we find
\begin{equation}
\label{diff1}
2m\cprod{\ve{\Omega}}{(\dot{\ve{\epsilon}}} - \ve{v}) =
\frac{1}{n}(\ve{\sigma} + \ve{\zeta}),
\end{equation}
expressing the balance of the Coriolis force on the left with the the
elastic stress and external perturbation.  The curl and divergence of this
equation yield,
\begin{equation}
 \label{curldiff1}
 \ve{\hat{z}}\ve{\nabla}\cdot(\dot{\ve{\epsilon}} - \ve{v}) -
 \frac{\partial}{\partial z}(\dot{\ve{\epsilon}} - \ve{v}) =
 \frac{1}{2\Omega nm}\cprod{\ve{\nabla}}{(\ve{\sigma} + \ve{\zeta})},
\end{equation}
and
\begin{equation}
 \label{divdiff1}
 (\cprod{\ve{\nabla}}{\dot{\ve{\epsilon}})_{z}} +
 2\Omega\ve{\nabla}\cdot\ve{\epsilon} =
 -\frac{1}{2\Omega nm}\ve{\nabla}\cdot(\ve{\sigma} + \ve{\zeta}).
\end{equation}

    For the rapidly rotating BEC, the energy to add a particle includes the
interaction energy, $\frac12gn^2$, the external potential, the quantum
pressure, and the elastic energy on account of the coupling of the particle and vortex
densities.  Thus,
\begin{eqnarray}
\mu_s &=& V_{\rm{eff}} + gn - \frac{\hbar^2}{2m\sqrt{n}}\nabla^2\sqrt{n} +
\frac{\delta\mathcal{E}_{\rm el}}{\delta n}
 \\
 &=& V_{\rm{eff}} + gn - \frac{\hbar^2}{2m\sqrt{n}}\nabla^2\sqrt{n} -
 \gamma n_{\rm v}^0\ve{\nabla}_{\perp}\cdot\ve {\epsilon},
\end{eqnarray}
where
\begin{equation}
 V_{\rm{eff}}(\ve{r}) = V(\ve{r}) - \frac{m}{2}\Omega^2 r_\perp^2
\end{equation}
is the trapping potential, $V$, plus the centrifugal potential.  For the
following calculations, we deal with a uniform particle density, which is the
case if the system is in a harmonic trap with a radial trap frequency equal to
the rotation rate, or if the superfluid is incompressible and in a cylindrical
container.

    We add perturbations of the density and velocity to linear
order.  From the Thomas-Fermi approximation, $\mu_s^{(0)}=V_{\rm{eff}} + gn_0$
in equilibrium, we have $\delta\mu_{s} = ms^2\delta n - \gamma n_{\rm v}^0\ve{\nabla}_{\perp}\cdot\ve {\epsilon}$, where $s=\sqrt{gn/m}$ is the speed of sound.  Taking the divergence of
Eq.~({\ref{accequ}}) and using the equation of continuity, we obtain
\begin{eqnarray}
 \left(-\frac{\partial^{2}}{\partial t^{2}} + s^{2}\nabla^{2} -
 \frac{\hbar^2}{4m^2}\nabla^4\right)\delta
 n &=& 2n\Omega(\cprod{\ve{\nabla}}{\dot{\ve{\epsilon}})_z} +
 \frac{\gamma n_{\rm v}^0n}{m}\nabla^2(\ve{\nabla}_{\perp}\cdot\ve{\epsilon})
 \label{densityelastic}
 \\
 &=&\left(\frac{\gamma n_{\rm v}^0n}{m}\nabla^2-
 4n\Omega^{2}\right)(\ve{\nabla}\cdot\ve{\epsilon}) -
 \frac{1}{m}\ve{\nabla}\cdot(\ve{\sigma} + \ve{\zeta}),
  \label{mainequ1}
\end{eqnarray}
where we note that $\ve{\nabla}_{\perp}\cdot\ve{\epsilon}=\ve{\nabla}\cdot\ve{\epsilon}$.
The density variations are coupled to the transverse motion of the
displacement field via the Corriolis force and to the longitudinal
displacement via the coupling of the particle and vortex densities.

    Taking the time derivative of Eq.~({\ref{curldiff1}}), using the curl of
Eq.~({\ref{stress}}) and Eq.~({\ref{divdiff1}}), and neglecting terms of 
the elastic moduli squared, we obtain
\begin{equation}
 \label{mainequ2}
 \left(-\frac{\partial^{2}}{\partial t^{2}} +
 \frac{2C_{2}}{nm}\nabla_{\bot}^{2}
 + \frac{\tau}{nm}\frac{\partial^{2}}{\partial
 z^{2}}\right)\ve{\nabla}\cdot\ve{\epsilon} =
 \frac{1}{n}\frac{\partial^{2}}{\partial t^{2}}\delta n +
 \frac{\gamma n_v}{(2\Omega nm)^2}\left(2C_2\nabla_{\perp}^2 + \tau\frac{\partial^2}{\partial z^2}\right)\nabla_{\perp}^2\delta n +
 \frac{\partial^2 }{\partial z \partial t}v_z.
\end{equation}
From the curl of the curl of Eq.~(\ref{velophase}), we have an expression for $v_z$:
\begin{equation}
\nabla^2 v_z = -\frac{1}{n}\frac{\partial^2}{\partial z \partial t}\delta n - 2\Omega\frac{\partial^2}{\partial z^2}(\cprod{\ve{\nabla}}{\ve{\epsilon}})_z
\end{equation}

\end{document}